\documentclass[reprint,showpacs,multibbl,amsmath,amssymb,prl]{revtex4-2}
\usepackage[english]{babel}
\usepackage[utf8]{inputenc}
\usepackage{graphicx,wasysym}
\usepackage{dcolumn}
\usepackage{bm}
\usepackage[colorinlistoftodos, color=green!40, prependcaption]{todonotes}
\usepackage[pdftex, pdftitle={Article}, pdfauthor={Author}]{hyperref}
\begin{document}
\title{Direct estimates of irreversibility from time series}

\author{Trevor GrandPre$^a$, Gianluca Teza$^b$, and William Bialek$^c$}
\affiliation{$^{a,c}$Joseph Henry Laboratories of Physics, $^{a,c}$Lewis--Sigler Institute for Integrative Genomics, $^a$Princeton Center for Theoretical Science, Princeton University, Princeton, NJ 08544 USA\\
and $^{b}$Max Planck Institute for the Physics of Complex Systems, 01187 Dresden DE }
 
\date{\today}

\begin{abstract}
The arrow of time can be quantified through the Kullback-Leibler divergence (\(D_{KL}\)) between the distributions of forward and reverse trajectories in a system. Many approaches to estimate this rely on specific models,   but the use of incorrect models can introduce   uncontrolled errors. Here, we describe a model-free method that uses trajectory data directly to estimate the evidence for irreversibility over finite windows of time.  To do this we build on previous work to identify and correct for errors that arise from limited sample size.   Importantly, our approach accurately recovers \(D_{KL} = 0\) in systems that adhere to detailed balance, and the correct nonzero \(D_{KL}\) for data generated by well understood models of nonequilibrium systems. We apply our method to trajectories of neural activity in the retina as it responds to naturalistic inputs, and find evidence of irreversibility in single neurons, emphasizing the non--Markovian character of these data.
These results open new avenues for investigating how the brain represents  the arrow of time.

\end{abstract}

\maketitle

The arrow of time is a salient fact about the world.  The second law of thermodynamics requires that this arrow exist, and the increase of entropy with time gives a quantitative measure of how different the world would look if time ran backwards.  Even if we make incomplete or coarse grained observations, so that we can't track all the heat flows and estimate the entropy production, we can see evidence for irreversibility in the visible trajectories of system state~\cite{martinez2019inferring,van2022thermodynamic, harunari2022learn, ehrich2021tightest, ghosal2022inferring, baiesi2024effective, ghosal2023entropy, teza2020exact, nitzan2023universal, skinner2021estimating,li2024measuring, pietzonka2024thermodynamic}.  This evidence has a precise information theoretic definition as the Kullback--Leibler divergence between the distribution of trajectories and their time reverse, and if we make complete observations this becomes equivalent to the thermodynamic entropy production~\cite{lebowitz1999gallavotti, seifert2005entropy, gaspard2004time, jarzynski2006rare, maes2003time, parrondo2009entropy}. This equivalence is central to many recent developments in non--equilibrium statistical physics \cite{Seifert2012, Esposito2010}. 

Most analyses of irreversibility focus on models.  As an example, if we have a system with states $\rm i$ and transitions are Markovian, then in steady state the rate of entropy production is
\begin{equation}
\sigma = {1\over 2} \sum_{\rm ij}\left( k_{\rm ij} P_{\rm j} - k_{\rm ji} P_{\rm i}\right)
\ln\left( {{k_{\rm ij} P_{\rm j}}\over{k_{\rm ji} P_{\rm i}}}\right) ,
\label{sigma1}
\end{equation}
where $k_{\rm ij}$ the probability per unit time for a transition ${\rm j} \rightarrow {\rm i}$, and $P_{\rm i}$ is the steady state probability of finding the system in state $\rm i$~\cite{schnakenberg1976network, van2015ensemble, mallick2009some, peliti2021stochastic, ziener2015entropy}.  While a complete microscopic description of a system must be Markovian, the limited set of variables that we can observe often have non--Markovian dynamics.  Nonetheless we would like to quantify the evidence that these variables provide about the arrow of time.

Imagine that we observe a system across a window of duration $T$, and call the observable trajectory $\gamma_T$. This trajectory is drawn from a distribution $P_T(\gamma_T)$.  Evidence for the arrow of time is contained in the fact that the probability of the time--reversed trajectory $\tilde\gamma_T$ is different.  This difference is measured by the Kullback--Leibler divergence $D_{KL}$, 
\begin{equation}
D_{KL}\left[ P_T(\gamma_T) || P_T(\tilde \gamma_T)\right] = \sum_{\gamma_T} P_T(\gamma_T) \ln\left[ {{P_T(\gamma_T) }\over{P_T(\tilde \gamma_T) }}\right] .
\end{equation}
We recall that $D_{KL}$ is positive semi--definite, reaches zero only if the two distributions are identical, and measures (colloquially) how certain we are that our observations came from one distribution rather than the other.  We emphasize that this does not depend on assumptions about the distribution of trajectories.  In the limit that the dynamics are Markovian we have
\begin{equation}
\lim_{T\rightarrow\infty} {1\over T} D_{KL} (T) = \sigma ,
\end{equation}
with $\sigma$ from Eq (\ref{sigma1}).
The $D_{KL}(T)$ metric can be applied to time series data to reveal the arrow of time~\cite{porporato2007irreversibility, roldan2010estimating, ro2022model, van2023time, kapustin2024utilizing,harunari2024uncovering, fritz2024entropy}. However, when working with finite datasets, systematic errors will arise. This paper aims to quantify these errors and develops a method to eliminate them, enabling more accurate estimates.

Even in a system with just two states, there are roughly one million possible trajectories spanning twenty time steps, and few experiments generate enough samples to see all of these.  More subtly, the nonlinearity of mapping between distributions and $D_{KL}$ means that random errors in the distribution become systematic errors in $D_{KL}$.  The same difficulties arise in estimating the entropy of a distribution, or the mutual information between different observables.  These problems were appreciated in early information theoretic analyses of human behavior~\cite{miller_55}, and received renewed attention in efforts to estimate the entropy and information content of neural responses~\cite{strong1998entropy, bialek2012biophysics}.    One approach, which we adopt here, is to search for expected systematic dependences of our estimates on the size of the data set, and extrapolate.

We start by discretizing the trajectory segment $\gamma_T$ into ``words'' $W$; we will denote by $\tilde W$ the word corresponding to the time reversed segment $\tilde\gamma_T$.   Discretization may be natural: the system may have a finite list of states, and observations may come only at discrete times.  If we impose some external discretization we will have to check how our results depend on this, and again we expect to find systematic dependences.    If the data we have to work with consist of $N$ independent samples of the words, and we observe $N_W$ examples of the word $W$, then the naive or frequentist estimate of the probability distribution is
\begin{equation}
P_{N} (W) = {{N_W}\over{N}} .
\end{equation}
We notice that this estimate is correct on average,
\begin{equation}
\langle P_{N} (W)\rangle  = P(W),
\end{equation}
and if the number of possible words is large the errors in the estimator obey
\begin{equation}
\langle \delta P_{N} (W) \delta P_{N} (W')\rangle  = { {\delta_{W,W'} } \over N} P(W).
\label{varP}
\end{equation}
As is well known, if we try to estimate the entropy by ``plugging in'' our estimate of the underlying distribution,
\begin{equation}
\hat S_N = -\sum_W P_{N} (W) \ln P_{N} (W) ,
\end{equation}
then we have a systematic bias \cite{miller_55,panzeri+treves_95}
\begin{equation}
\langle \hat S_N \rangle  = -\sum_W P  (W) \ln P  (W)  - {A\over N} - { B \over{N^2}} + \cdots .
\label{Sbias}
\end{equation}
If we can trust that all $N$ samples really are independent, then $A = \Omega/2$, where $\Omega$ is the number of possible words, and this is independent of the distribution.  We can obtain accurate estimates of the true entropy by identifying this dependence on $N$ and extrapolating $N\rightarrow \infty$. This approach is justified by the fact that, as word length increases, each word approaches independence, since correlations between consecutive words diminish subextensively with longer word lengths.

\begin{figure}[b]
\includegraphics[width=0.95\linewidth]{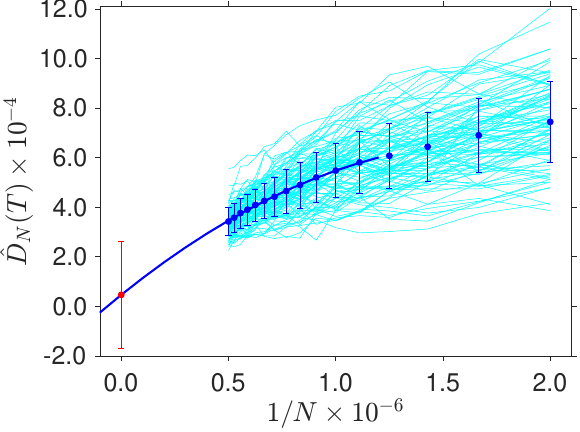} 
\caption{Recovering zero in a two--state Markov system.  Plug--in estimates $\hat D_N (T)$ vs the (inverse) number of samples $N$ (cyan curves), with mean and standard deviation across these results (blue points with error bars); here $T = K\Delta\tau$, where $\Delta\tau=0.01\,{\rm s}$ and $K=10$.  Each $\hat D_N (T)$  is fit to Eq (\ref{DKLvsN}) and we show the mean (blue line), as well as the mean and standard deviation of the extrapolation $N\rightarrow\infty$ (red point with error bar).\label{zero}}
\end{figure}

In the case of $D_{KL}$ things are a bit more complicated.  The path to Eq (\ref{Sbias}) is to expand $\hat S_N$ in $\delta P$ and then use Eq (\ref{varP}) to compute the average.   The plug--in estimator of $D_{KL}(T)$ is
\begin{equation}
\hat D_{N} (T) = \sum_W P_N(W) \ln\left[ {{  P_N(W)}\over{P_N(\tilde W)}}\right] ,
\end{equation}
and following the same path as for the entropy gives
\begin{eqnarray}
\langle \hat D_{N} (T) \rangle &=& D_{KL}(T) + {A\over {N}}   + {B\over {N^2}} + \cdots ,
\label{DKLvsN}\\
A&=& {1\over 2}\left[\Omega' + \sum_W{{P(W)}\over{P(\tilde W)}} \right] ,
\end{eqnarray}
where $\Omega'$ is the number of words that are distinguishable from their time-reversed version. We see that systematic errors exhibit the same form as for the entropy, but even the term   $\sim 1/N$  is not universal. Nonetheless, we can look for this systematic behavior and extrapolate.

Note that we expect a systematic {\em over}--estimate of $D_{KL}$, essentially because in finite random samples we observe $W$ and $\tilde W$ different numbers of times even if the underlying dynamics are time reversal invariant.  This is in contrast to the entropy, where the plug--in estimator has the opposite bias, but similar to what happens when we estimate mutual information.

To see how this works let's start with a Markovian two state system, for which we know the correct answer is $\sigma = 0$.  We choose the transition probabilities per unit time to be $k_{12} = k_{21} = 1\,{\rm s}^{-1}$, and discretize time into bins of size $\Delta\tau = 0.01 \,{\rm s}$.  The trajectories now are binary strings, and the words $W$ are binary words; if we look at windows of duration $T = K\Delta\tau$ then $\Omega = 2^K$. In Figure \ref{zero}, we follow the trajectory of $\hat{D}_N(T)$, generated through discretized continuous-time Monte Carlo simulations, as successive samples of words with length $K = 10$ are added. We observe small nonzero values for the estimate of $D_{KL}$ even at $N \sim 10^6$, and repeated sampling with one million samples shows that these values are statistically significant. However, by allowing $N$ to vary, we observe behavior consistent with Eq.~\(\ref{DKLvsN}\), which enables us to extrapolate each trajectory. The mean result is within one standard deviation of zero, as expected.

For a more positive example we consider a three state system, which we can think of as a cycle. The probability per unit time for clockwise transitions is chosen as 
\begin{equation}
k_{12} = k_{23} = k_{31} = k_{CW} = 1 \,{\rm s}^{-1},
\end{equation}
and for counterclockwise transitions 
\begin{equation}
k_{13} = k_{32} = k_{21} = k_{CCW} = 10\,{\rm s}^{-1}.
\end{equation}
This system violates detailed balance and from Eq (\ref{sigma1}) we have $\sigma = (10 -1)\ln(10) \approx 20.7$.  We discretize time into bins of either $\Delta\tau = 0.1$  or $0.05\,{\rm s}$; words $W$ now are ternary words, and can always be labelled by the corresponding base ten number.  As before we simulate long enough to generate a few million words, and extrapolate using Eq (\ref{DKLvsN}). The result still depends on the duration of the window $T$.

\begin{figure}
\includegraphics[width=\linewidth]{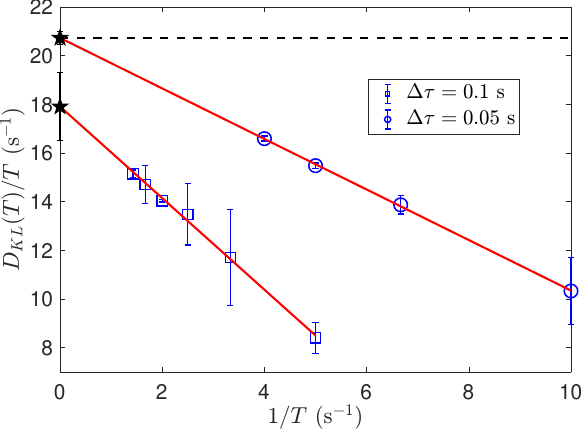}
\caption{Irreversibility in a three state system. Extrapolated $D_{KL}(T)$ from Eq (\ref{DKLvsN}), as a function of (inverse) window duration $T=K \Delta \tau $ (points with error bars), with different temporal bin sizes $\Delta\tau$ (legend), and word lengths $K=3-12$.   Lines are from Eq (\ref{DtoSigma}), showing convergence to a rate $\sigma$, which agrees with the analytic estimate from Eq (\ref{sigma1}) if $\Delta\tau$ is sufficiently small. \label{3state}}
\end{figure}

In a system with finite correlation times we expect that the subextensive contributions to entropy--like quantities will be constant, so that at large $T$ we have
\begin{equation}
D_{KL}(T) \rightarrow \sigma T + \sigma_1 ,
\label{DtoSigma}
\end{equation}
and this is what we see in Fig \ref{3state}.  Not only do we see a well defined rate $\sigma$, but the value agrees with the analytic prediction from Eq (\ref{sigma1}), at least if the time bins are sufficiently small.  It is perhaps not surprising that we need $\Delta\tau \times \max(k) < 1$ to get completely reliable results.

\begin{figure}[b]
\includegraphics[width=0.87\linewidth]{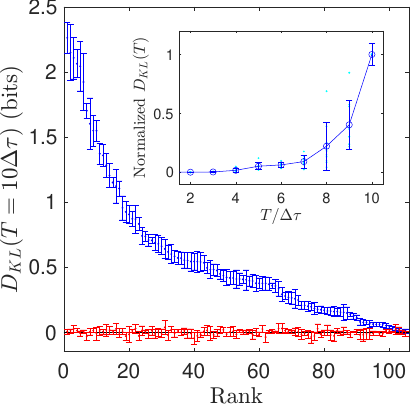} 
\caption{The arrow of time in the activity of single neurons.  Main figure shows our best estimates of $D_{KL}(T = 200\,{\rm ms})$ for individual retinal ganglion cells (blue points with error bars), based on the experiments described in Ref~\cite{tkavcik2014searching}.  As a check we shuffle the data for each cell along the time axis, and verify that our estimation procedure gives $D_{KL} = 0$ within error bars (red points with error bars).  Inset shows the dependence of $D_{KL}(T)$ on $T/\Delta \tau$ normalized by $D_{KL}(T=10\Delta\tau)$ and averaged over the ten neurons with the largest $D_{KL}$.  Error bars include both statistical errors and the standard deviation of means across these ten cells.
\label{neurons}}
\end{figure}

Having calibrated our method, we now turn to experiments on the neural representation of the arrow of time.  Specifically we consider experiments that monitor the output neurons of the salamander retina (retinal ganglion cells) as it responds to naturalistic grayscale movies~\cite{tkavcik2014searching}.  These experiments resolve the sequences of  individual action potentials (spikes) from 100+ neurons in a small patch, and it is conventional to discretize time into bins of duration $\Delta\tau = 20~{\rm ms}$; this is chosen small enough to insure that we almost never see two spikes in one bin but long enough to capture the dominant correlations between neurons.  In Fig \ref{neurons} we see the result of estimating $D_{KL}(T = 200\,{\rm ms})$ for individual neurons.  An important check on our control over errors from finite sample size  is that when we shuffle the data, breaking temporal correlations, we find $D_{KL} = 0$ within error bars.

Recall that for a binary sequence we have $D_{KL} = 0$ if the dynamics are Markovian.  The fact that we find almost all neurons to have nonzero evidence for irreversibility implies strongly non--Markovian behavior, which perhaps is not surprising.  But in order to ``feel'' the non--Markovian structure one needs to look over multiple time steps.  As we see in the inset to Fig \ref{neurons}, this leads to a super--linear growth of $D_{KL}(T)$ as the time window $T$ of our observations increases; this is happening on the $40 < T < 200\,{\rm ms}$ scale of relevance to perception and behavior.  Presumably there is a crossover to $D_{KL}(T) \sim \sigma T$ at very large $T$, but this is beyond the scale on which we can make reliable model--independent estimates from experiments of reasonable duration.

In assessing the evidence for the arrow of time from experiment, we have a choice between making models of the underlying processes and trying to use the data directly.  But wrong models introduce uncontrolled errors.  Our analysis of single neurons provides an extreme example of this problem. These signals are naturally discretized into sequences of spiking and silence, and Markov models of such binary sequences obey detailed balance for all parameter values.  By looking directly at the data we see, in contrast, that single neurons in the retina provide substantial evidence for the arrow of time as they encode naturalistic inputs.  While we usually expect that entropy and related quantities approach extensive behavior from above, this is not true for the irreversibility $D_{KL}(T)$, which we find grows supralinearly on perceptually relevant time scales.  Successive small time windows combine evidence for the arrow of time in a synergistic manner.  It would be interesting to find minimal models that reproduce these unexpected behaviors.

Our ability to draw conclusions from data directly depends on correcting for the systematic errors that arise from finite sample size.  While there is no magic, there is a regime in which systematic errors are significant but can be removed by recognizing the predicted dependence on sample size and extrapolating.   This follows a strategy widely used in the analysis of information flow in neural coding.  Related work uses lower bounds on the entropy from coincidence probabilities \cite{ma1981}, upper bounds from the maximum entropy construction \cite{meshulam+bialek2024}, and Bayesian estimators that implement a uniform prior on the entropy of the system \cite{nsb,nemenman+al2004}.  More recent work, in the same spirit, uses bounds from the thermodynamic uncertainty relation to improve direct estimates of the entropy production rate \cite{li2019}.  

There are interesting open questions about how to better use analytic bounds to improve direct estimates of $D_{KL}$ from data.  As an example, by analogy with maximum entropy methods one could ask how to construct models that have the minimal $D_{KL}$ consistent with measured correlation functions.  These ideas could be applied more broadly to estimate the many characteristics of non-equilibrium systems that underpin thermodynamic equalities~\cite{jarzynski1997nonequilibrium,suarez2012phase}, fluctuation theorems~\cite{crooks1999entropy,stella2023anomalous,coghi2023convergence,collin2005verification}, and large deviation functions~\cite{levien2020large,hidalgo2017,ray2018,grandpre2021,teza2024universal,stella2023universal,teza2020rate,angeli2019}.

 fundamental feature of all direct methods is a tradeoff between the size of the sample that we consider and the duration of the trajectories that we can observe.  One notable example is in the estimation of population growth rates from single-cell lineages, which relies on the evaluation of the large deviation function~\cite{levien2020large}. It was found that there was a goldilocks principle between the number of samples and the lineage length with a delicate balance of these two errors which gives the best estimate of the population growth rate. 

In summary, we have argued that direct estimates of the evidence for the arrow of time can be effective in places where model--based estimators are not.  These methods should help us explore the representation of irreversibility in neural activity and in other systems which lack observable thermodynamic fluxes.

\begin{acknowledgements}
This work was supported in part by the National Science Foundation, through the Center for the Physics of Biological Function (PHY--1734030), by  the Schmidt Science Fellowship (TG), and by fellowships from the Simons Foundation and the  John Simon Guggenheim Memorial Foundation (WB).
\end{acknowledgements}

\end{document}